\documentclass[secnumarabic,amssymb, superscriptaddress, nobibnotes, aps,prl,twocolumn,longbibliography]{revtex4-1}

\usepackage{amsmath}
\usepackage{graphicx, nicefrac}
\usepackage{mathtools}
\usepackage{xcolor}

\newcommand{\Fex}[1]{Fe$_{#1}$NbS$_2$}

\begin{document}
\title{Antiferromagnetic Switching Driven by the Collective Dynamics of a Coexisting Spin Glass}

\newcommand{\UCB}{Department of Physics, University of California, Berkeley, CA 94720, USA}
\newcommand{\LBL}{Materials Sciences Division, Lawrence Berkeley National Laboratory, Berkeley, California, 94720, USA}

\newcommand{\UCBC}{Department of Chemistry, University of California, Berkeley, California 94720, USA}
 
\newcommand{\NHMFL}{National High Magnetic Field Laboratory, Tallahassee, Florida 32310, USA}

\newcommand{\UCBM}{Department of Materials Science and Engineering, University of California, Berkeley, California 94720, USA}

\author{Eran Maniv}
\affiliation{\UCB}
\affiliation{\LBL}

\author{Nityan Nair}
\affiliation{\UCB}
\affiliation{\LBL}

\author{Shannon C. Haley}
\affiliation{\UCB}
\affiliation{\LBL}

\author{Spencer Doyle}
\affiliation{\UCB}

\author{Caolan John}
\affiliation{\UCB}

\author{Stefano Cabrini}
\affiliation{Molecular Foundry, Lawrence Berkeley National Laboratory, Berkeley, California 94720, USA}

\author{Ariel Maniv}
\affiliation{NRCN, P.O. Box 9001, Beer Sheva, 84190, Israel}
\affiliation{\NHMFL}

\author{Sanath K. Ramakrishna}
\affiliation{\NHMFL}

\author{Yun-Long Tang}
\affiliation{\UCB}
\affiliation{\UCBM}
\affiliation{\LBL}

\author{Peter Ercius}
\affiliation{Molecular Foundry, Lawrence Berkeley National Laboratory, Berkeley, California 94720, USA}

\author{Ramamoorthy Ramesh}
\affiliation{\UCB}
\affiliation{\UCBM}
\affiliation{\LBL}

\author{Yaroslav Tserkovnyak}
\affiliation{Department of Physics and Astronomy, University of California Los Angeles, California, USA}

\author{Arneil P. Reyes}
\affiliation{\NHMFL}

\author{James Analytis}
\affiliation{\UCB}
\affiliation{\LBL}

\date{\today}

\begin{abstract}
The theory behind the electrical switching of antiferromagnets is premised on the existence of a well defined broken symmetry state that can be rotated to encode information. A spin glass is in many ways the antithesis of this state, characterized by an ergodic landscape of nearly degenerate magnetic configurations, choosing to freeze into a distribution of these in a manner that is seemingly bereft of information. In this study, we show that the coexistence of spin glass and antiferromagnetic order allows a novel mechanism to facilitate the switching of the antiferromagnet \Fex{1/3+\delta}, which is rooted in the electrically-stimulated collective winding of the spin glass. The local texture of the spin glass opens an anisotropic channel of interaction that can be used to rotate the equilibrium orientation of the antiferromagnetic state. The use of a spin glass' collective dynamics to electrically manipulate antiferromagnetic spin textures has never been applied before, opening the field of antiferromagnetic spintronics to many more material platforms with complex magnetic textures.
\end{abstract}

\maketitle

There are a handful of material systems whose antiferromagnetic (AFM) spin texture can be electrically manipulated or `switched'.\cite{wadley2016electrical,bodnar2018writing,nair2019electrical} The mechanism is generally explained with the same underlying physics; an applied current induces a spin-polarization due to a combination of inversion asymmetry and spin-orbit coupling, that then transfers angular momentum into the system, exerting a `spin-orbit torque' that is able to manipulate the magnetic domains of the ordered state. Technically, this is referred to as a rotation of the N\'eel vector, which defines the orientation of a domain.\cite{gomonay2010spin,vzelezny2014relativistic,gomonay2017concepts} 
This in turn rotates the principal axes of the conductivity of the material, providing a switch between high and low resistance states along perpendicular directions. The `high/low' contrast of the switching is determined not only by the efficacy with which angular momentum can be transferred to the magnetic lattice by the applied current, but also by the degree of conductivity anisotropy within a domain. 

\begin{figure*}
\centering
\includegraphics[width=0.75\linewidth]{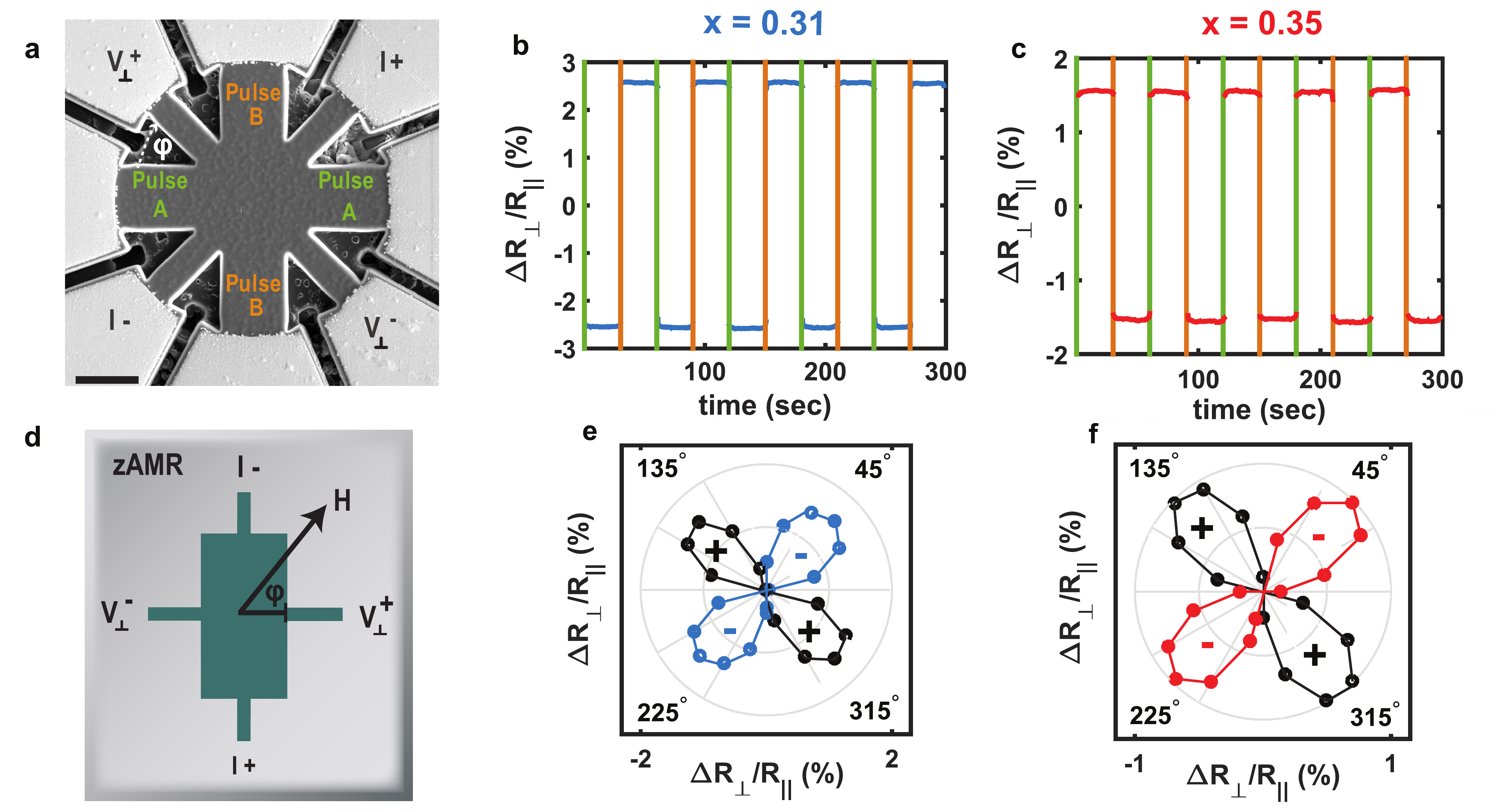}
\caption{{\bf The Electrical Switching and zAMR Effects.} (a) Scanning electron microscope image of a standard focused ion beam device fabricated on a $x = 0.35$ crystal. The device was fabricated in the a-b plane of the \Fex{x} crystal. The AC configuration measurement (read-out) and the DC current pulses (write-in) are marked on top of the image. All switching devices presented in the main text were measured by this configuration (switching 45$^{\circ}$). The scale-bar size is 10 $\mu$m. (b),(c) The electrical current switching response measured at 2 K for $x = 0.31$ and $x = 0.35$ intercalations respectively. A sequence of five A-B pulses was applied with a 30 seconds delay between pulses. The DC pulse amplitude was set at 54 kA/cm$^{2}$ (63 kA/cm$^{2}$) for the $x = 0.31$ (0.35) intercalation for a duration of 10 ms. The difference between the two switching states is presented in the form of the background subtracted transverse resistance ($R_{\perp}$-$R_{b}$) normalized by the longitudinal resistance ($R_{//}$). The A and B pulses are plotted as green and orange lines respectively. For both intercalations a stable switching amplitude is demonstrated with an opposite response to the pulse sequence. (d) Illustration of a typical AC contact configuration measured to probe the zAMR effect. (e),(f) Angle dependent polar plots of the zAMR effect measured at 2 K for $x = 0.31$ and $x = 0.35$ intercalations respectively. The samples were cooled in a magnetic field of 9 T, at various in-plane angles ($\phi$). Subsequently reaching base temperature the magnetic field was turned off and the transverse resistance ($R_{\perp}$) was measured. The zAMR was calculated by subtracting an average background and normalizing to the longitudinal resistance ($R_{//}$) measured simultaneously. Black circles represent a positive zAMR response, while blue (red) circles represent a negative response for $x = 0.31$ ($x = 0.35$). A similar zAMR response between the two compositions is observed.}
\label{fig:Intro}
\end{figure*}

The system \Fex{1/3} lacks inversion symmetry, so the spin orbit coupling will cause a partial spin polarization of a current pulse.\cite{baltz2018antiferromagnetic,edelstein1990spin}
In Figure \ref{fig:Intro}, we show the basic 8-terminal device configuration fabricated from single crystals of \Fex{1/3+\delta} (see Methods for synthesis and Focused Ion Beam microstructuring). In panels \ref{fig:Intro}b and c, we illustrate `pulse trains’ for two off-stoichiometry compositions, $x=0.31, 0.35$. Successive vertical and horizontal pulses take the system from high to low resistance states just as in other switchable AFMs, but with two key differences.\cite{wadley2016electrical,bodnar2018writing} First, there is single pulse saturation of the signal, independent of the current density used, with no detectable relaxation to some intermediate resistance. Second, the pulse widths and current densities used are orders of magnitude lower than other systems, typically $\sim$10$^6$ A/cm$^2$, whereas we observe switching at $\sim$ 10$^4$A/cm$^2$.\cite{wadley2016electrical,bodnar2018writing}
Both these properties have obvious advantages technologically, but it is far from understood why these occur in this material but not in others. In this study, we show that the answer to this question is, surprisingly, disorder. Disorder spawns a spin glass with its own collective dynamics,\cite{halperin1977hydrodynamic} capable of transferring angular momentum to the coexisting AFM.\cite{ochoa2018spin} There is one curious observation that distinguishes the compositions shown in Figure \ref{fig:Intro}; the electrical switching has a different phase for $x=0.31$ than for $x=0.35$. The N\'eel vector is being oriented in perpendicular directions in the dilute and excess iron compounds under the same direction of the current pulse. We shall revisit this observation later.

\begin{figure*}
\centering
\includegraphics[width=\linewidth]{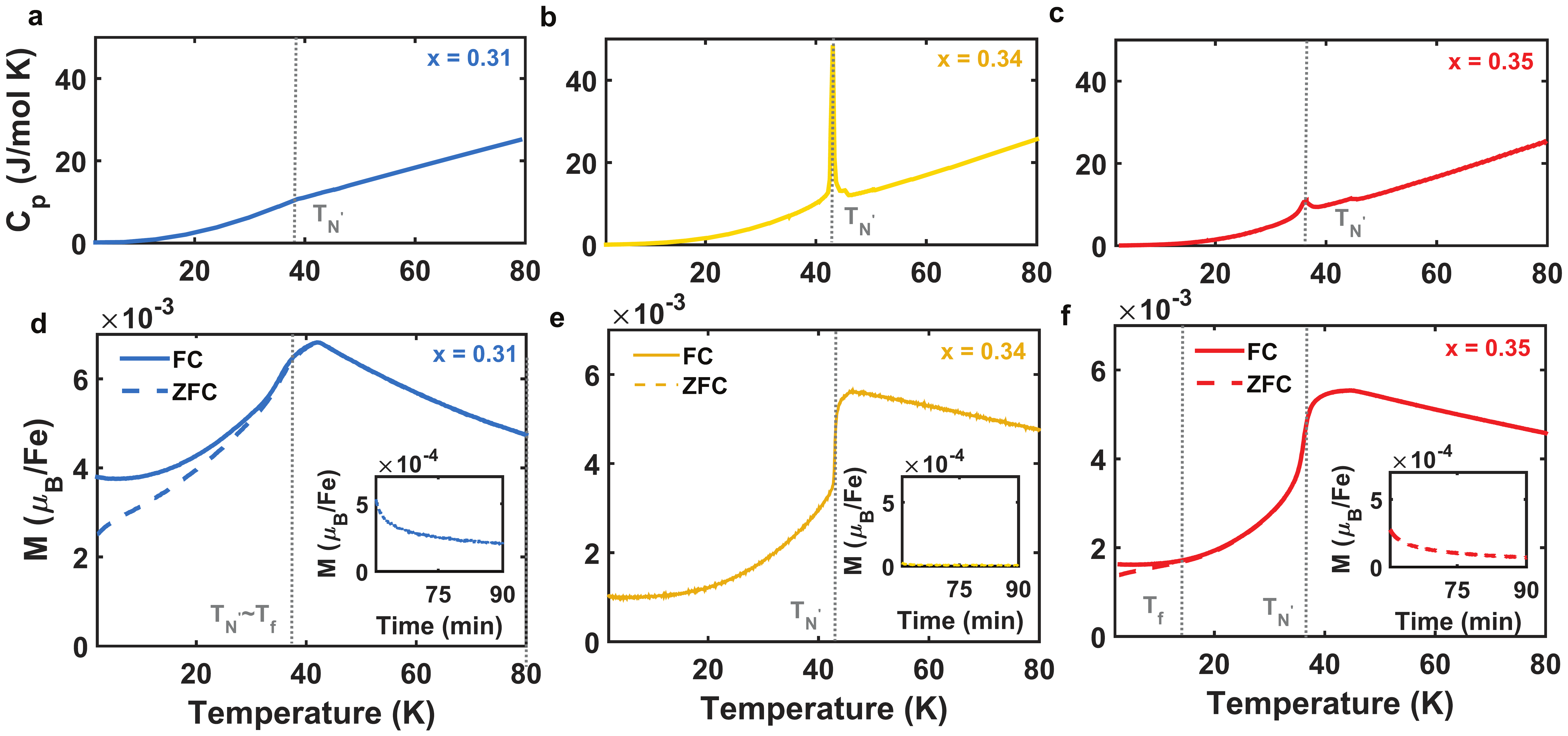}
\caption{{\bf Thermodynamic  Characterization of Fe$_x$NbS$_2$ for x = 0.31, 0.34 and 0.35.} (a)-(c) Heat capacity versus temperature curves for all intercalation values measured without an applied magnetic field. The dashed line marks the lower AFM transition (T$_{N^{'}}$), associated with an in-plane order. The vanishing heat capacity anomalies away from x = \nicefrac{1}{3} are correlated with the entrance of a glassy state. (d)-(f) 1000 Oe magnetization versus temperature curves for each intercalation value: both the field cool (FC - solid lines) and zero field cool (ZFC - dashed lines) curves are presented. The magnetic field was applied in the perpendicular direction with respect to the NbS$_{2}$ layers (c-axis). The divergence of the FC and ZFC curves demonstrates the onset of glassy behavior, i.e. the spin glass freezing temperature (T$_{f}$). Insets: Relaxation of the magnetization for x = 0.31, 0.34 and 0.35 intercalations at 5 K after a 1 T field was applied for 1 hour. The measurement is presented after the magnetic field was removed. Further analysis related to glassy dynamics is presented in the Supplemental Information.}
\label{fig:thermo}
\end{figure*}

Although the N\'eel order is mostly oriented out of plane,\cite{van1971magnetic} the high and low resistance states are likely associated with the re-orientation of a small in-plane component. This appears to be associated with an order parameter that causes a second transition at a lower temperature $T_{N'}$, appearing as a larger heat capacity anomaly (see Figure \ref{fig:thermo}b). In-plane studies of the nuclear magnetic resonance confirm the presence of an in-plane component to the AFM order (See Supplemental Information). This is also confirmed by measurements of a zero-field anisotropic magnetoresistance (zAMR), shown in Figure \ref{fig:Intro}d-f, where cooling in an in-plane field permanently re-orients the in-plane N\'eel vector. The zAMR consistently onsets at $T_{N'}$, consonant with the association of this transition with the in-plane canting of the moments. The zAMR in principle reflects the same conductivity as the `high/low' states of switching: a convolution of the average orientation of the in-plane N\'eel vector and the degree of anisotropy in the domain conductivity.\cite{kriegner2016multiple,vzelezny2018spin} 

\begin{figure*}
\includegraphics[width=\linewidth]{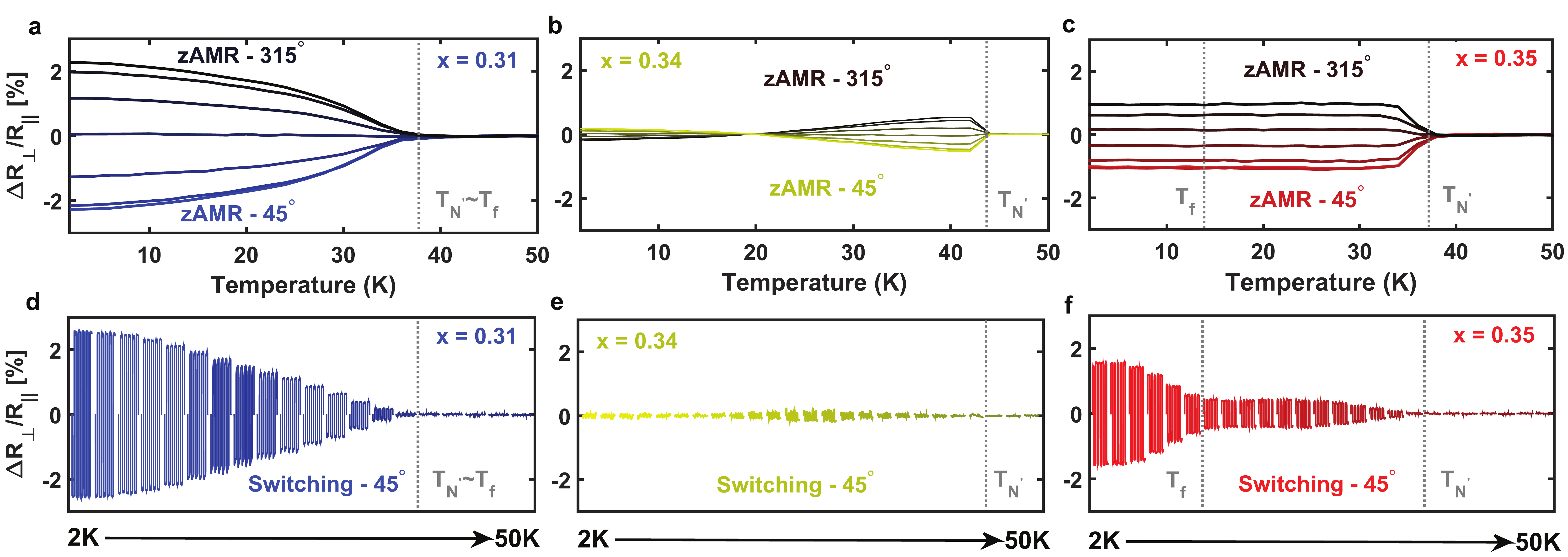}
\caption{{\bf Electrical Switching and zAMR Temperature Dependencies.} The temperature dependent zAMR measurements of $x = 0.31$ (a), $x = 0.34$ (b) and $x = 0.35$ (c) intercalations are plotted. The FC angle window (zAMR 315$^{\circ}$ to 45$^{\circ}$), for each intercalation, is presented along the correlated curves. The zAMR onset corresponds to the AFM transition temperature (T$_{N^{'}}$), with no distinct response to the spin glass freezing temperature (T$_{f}$). In the lower panels we plot the electrical current switching response of $x$ = 0.31 (d), 0.34 (e) and 0.35 (f) intercalations as a function of temperature. All switching devices were probed in the ``switching 45$^{\circ}$" configuration with a 100 $\mu$A (0.1 - 0.3 kA/cm$^{2}$) AC read-out current. All three plots are scaled similarly for comparison. To achieve switching, current densities of the order of 40 - 80 kA/cm$^{2}$ and pulse widths of the order of 1 - 10 ms were applied. For more information regarding pulse amplitude and width dependence see Supplemental Information. For $x = 0.35$ (f) an enhanced switching response appears at the same temperature where the spin glass state starts to freeze (T$_{f}$).}
\label{fig:TempDep}
\end{figure*}

In Figures \ref{fig:thermo} and \ref{fig:TempDep} we compare the temperature dependent properties of the $x=0.31$ and $x=0.35$ compositions to stoichiometric samples near $x=1/3$. The thermodynamic properties are straightforward; the $x=0.34$ system has the largest and sharpest heat capacity and magnetic anomaly at the N\'eel transition $T_{N'}$, broadening significantly at compositions off stoichiometry.\cite{tsuji1999heat,yamamura2004heat} The magnetic susceptibility also shows that spin glass dynamics are only present off stoichiometry, manifested as slow relaxation of the magnetization (see insets Figure \ref{fig:thermo}d-f). The sensitivity of the spin glass to its history causes the field-cooled and zero-field cooled curves to separate at a characteristic freezing temperature $T_f$, an effect absent in $x=0.34$, as observed in Figure \ref{fig:thermo}d and f.\cite{Doi1991} Other effects characteristic of glassy dynamics including aging and hysteresis about zero field are also present at these compositions (See Supplemental Information).\cite{mydosh2014spin} The temperature dependence of the zAMR signal (Figure \ref{fig:TempDep}a-c) passes through the freezing of the spin glass with impunity, reflecting only the smooth growth of the AFM order parameter as the temperature is lowered. Because the spin glass is invisible to the zAMR, it provides a reference point for the AFM response, untethered to the spin glass.

\begin{figure*}
\centering
\includegraphics[width=\linewidth]{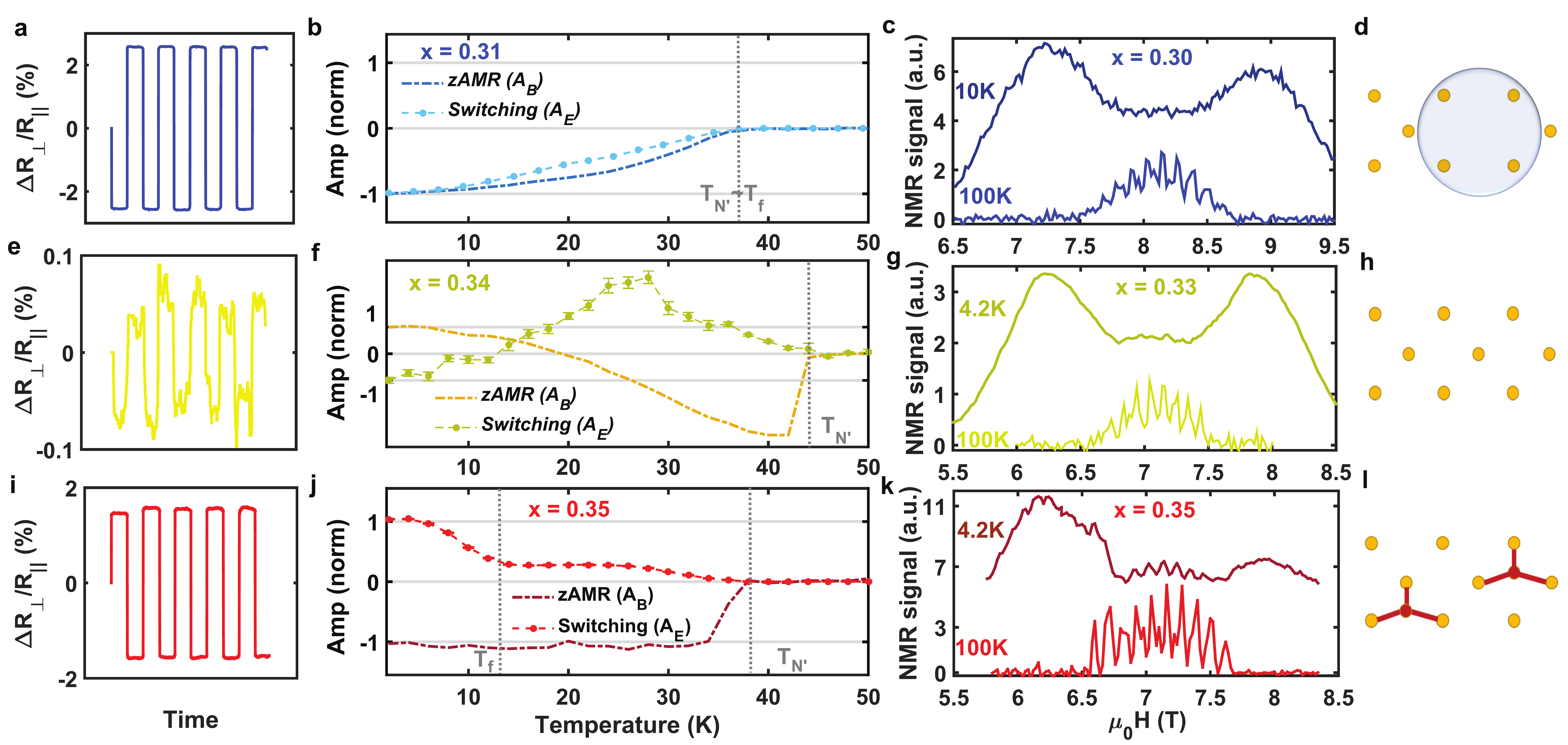}
\caption{{\bf Summary and NMR Measurements.} The low temperature switching response for $x$ = 0.31 (a), 0.34 (e) and 0.35 (i) illustrates the stability and enhanced switching amplitude while departing $x = \nicefrac{1}{3}$ value. Comparison between the electrical switching (full circles) and zAMR (dashed lines) temperature dependence for $x$ = 0.31 (b), 0.34 (f) and 0.35 (j) is presented. The switching amplitude (A$_{E}$) is defined as the difference in the relative resistivity change $\Delta R_{\perp}/R_{//}$ between A and B pulses. The zAMR amplitude (A$_{B}$) is defined as the difference in the relative resistivity change $\Delta R_{\perp}/R_{//}$ between $\phi = 45^{\circ}$ and $\phi = 135^{\circ}$. Both switching and zAMR are normalized by their low temperature absolute value amplitude for comparison.
The sign of the switching amplitude, low-high (sgn(A$_{E}$)$=-1$) or high-low (sgn(A$_{E}$)$=+1$) can be seen to depend on intercalation x and, for the sample near $x = \nicefrac{1}{3}$, on temperature too.
The transition from a correlated switching-zAMR temperature dependence (b; $x = 0.31$) to an anti-correlated temperature dependence (j; $x = 0.35$) is clearly elucidated in this intercalation regime.
Field-swept $^{93}$Nb NMR spectra at 85 MHz (c; $x = 0.30$), 74.5 MHz (g; $x = 0.33$) and 74.5 MHz (k; $x = 0.35$) measured above and below the AFM transition. All sweeps were performed after cooling the samples in a perpendicular magnetic field (c-axis). In the paramagnetic regime the field sweeps show the $^{93}$Nb quadrupolar splitting for all intercalations. At low temperatures two broad peaks indicative of a long-range AFM order emerge. For $x = 0.30$ (c) and $x = 0.35$ (k) samples an asymmetry component appears on-top of the low temperature peaks. Additionally, for the $x = 0.35$ sample the paramagnetic Nb peaks are visible at low temperatures, indicating the magnetic lattice acquires an additional spin species. An illustration of the iron lattice is presented for the relevant regimes: vacancies (d), stoichiometric (h) and interstitials (l).}
\label{fig:summupNMR}
\end{figure*}

The temperature dependence of the electrical switching of the N\'eel vector offers a surprising contrast to the thermodynamic response: it is strongly suppressed for compounds near stoichiometry and in all respects enhanced when the spin glass is present (Figure~\ref{fig:TempDep}d-f). The interplay of spin glass and AFM order is especially pronounced in the $x=0.35$ composition. Notice that even though the N\'eel canting transition occurs at $T_{N'}\sim 37$K, there is a large enhancement of the switching at $\sim15$K (Figure~\ref{fig:TempDep}F). There is no (re)ordering phase transition in this range, but it is exactly the temperature where the spin glass freezes, $T_f$. At $x=0.31$, $T_f$ and $T_{N'}$ coincide, so that the switching simply follows the growth of the AFM order parameter (Figure~\ref{fig:TempDep}d). The data in Figure~\ref{fig:TempDep} also illustrates another important point: the switching of stoichiometric compositions is not only significantly smaller than in the diluted or excess case, but also significantly less {\it stable}. As can be observed in the enlargement at low temperatures (Figure \ref{fig:summupNMR}a,e,i), the signal for the $x = 0.34$ intercalation varies from pulse to pulse by up to 20$\%$, in comparison to intercalation where the spin glass is present, where the signal is stable within 0.5$\%$. The coexistence of the spin glass greatly increases the efficacy of the spin current in transferring angular momentum to the system, leading to an enhancement in both amplitude and stability of the switching. The switching of \Fex{1/3+\delta} therefore depends on the interplay of the responses of two coupled order parameters, the AFM and the spin glass.

To better understand the mechanism of this interplay we study the local environment of magnetic moments with nuclear magnetic resonance (NMR), shown in Figure \ref{fig:summupNMR}c,g,k. The iron exchange field is studied via its effect on the $^{93}$Nb lattice (with nuclear spin I = 9/2, $\gamma$ = 10.405 MHz/T). In the paramagnetic state at temperatures $T>T_{N'}$, the spectra exhibit a broad peak with quadrupolar splitting originating from two Nb unit cell sites. Below $T_{N'}$ the system splits into a double-peak structure symmetric about the paramagnetic center. This is a signature of AFM order, with the two peaks originating from the two sublattices where the hyperfine field ($\sim$1 T) adds to, and subtracts from, the applied magnetic field.\cite{buttgen2012high} Even though the peak structure is broadly the same at all compositions (reflecting a similar AFM order for all $x$), on cooling in an out-of-plane field an important difference emerges for compositions that are off-stoichiometry; the peaks become asymmetric in magnitude. This strongly suggests that the spin glass exerts an exchange field on the AFM lattice, causing spin-flips that align with one sublattice. Moreover, since it is always the left-most peak that is enhanced, the exchange coupling $J$ of the spin glass to it's AFM neighbors is likely ferromagnetic (FM) ($J>0$) for both dilute and excess compositions. This provides strong evidence for the exchange coupling between the spin glass and the AFM order parameters.

\section*{DISCUSSION}
The above data establish three facts about the nature of the switching in \Fex{1/3+\delta} that distinguishes this system from any known counterpart. Firstly, in order for a spin polarized current to rotate the N\'eel state of the system with any efficacy, the system needs to be disordered, the spin glass must be present. This distinguishes the N\'eel rotation due to switching from that caused by applied field in the zAMR, which is not affected by the spin glass. Second, there is a local exchange interaction that couples the spin glass to the AFM lattice. Third, the freezing of the spin glass has the dual effect of enhancing the current induced rotation of the N\'eel state, and then pinning it to create a large and stable `high/low' contrast.  

In order to understand what is special about the coexistence of the spin glass and the AFM, it is notable that in most other examples of switchable AFMs the switching does not saturate with a single pulse, but rather shows saw-tooth behavior.\cite{wadley2016electrical,bodnar2018writing} This is explained by assuming the majority of the response is driven by shifts in the AFM domain boundaries\cite{li2015activation,wadley2018current}, the motion of which will change the average direction of the N\'eel vector. Domain boundaries generally form near structural defects in the material, and so the switching depends on the ability of these defects to de-pin and move through the AFM lattice. Moreover, increased disorder will tend to make the electron scattering more isotropic, which will usually reduce the `high/low' contrast of the switching.\cite{baltz2018antiferromagnetic} The present situation appears at first sight to be in the opposite limit; not only do defects need to be present, but the associated spin glass needs to be {\it frozen} ($T<T_f$) for the switching to become pronounced and stable. This implies that the freezing of the spin glass opens a new channel for the transfer of angular momentum, one that leverages the local stiffness of the spin glass itself. 

The connection to stiffness suggests that the collective motion of the spin glass is transferring the spin torque. A related concept to this has been discussed in the context of spin hydrodynamics of insulating correlated spin glasses.\cite{ochoa2018spin} The essential idea can be understood by describing the spin glass as a rotation-matrix valued order parameter describing an overall orientation of a volume of mutually disordered but frozen spins. (This follows the treatment introduced by Halperin and Saslow\cite{halperin1977hydrodynamic} where this object can be connected to  the Edwards-Anderson order parameter of a spin glass). A spin accumulation will generate a collective winding of this volume of spins, which is completely analogous to spin torques across interfaces,\cite{tserkovnyak2017generalized} with a precession frequency that depends on the ratio of the relevant spin-mixing conductance and the Gilbert damping of the spin glass. In principle, if there is a coexisting AFM, as in the present case, collective motion imparts spin torque on the N\'eel vector. Although the magnetic disorder and associated local anisotropies mean that spin is not conserved locally, the spin texture is topologically constrained by its spatiotemporal winding characteristics, with the net spin being the generator of the winding.\cite{ochoa2018spin} The winding dynamics thus appears as a net non-equilibrium spin, amplifying the spin transfer from the electronic spin accumulation of the current pulse.

This mechanism may also help explain another unusual feature in the switching of \Fex{1/3+\delta}, briefly introduced in Figure \ref{fig:Intro}: the sign of $\delta$ determines which direction the N\'eel vector is rotated during the current pulse. Dilute compositions ($\delta<0$), where the defects are predominantly vacancies, switch in the opposite direction to excess compositions ($\delta>0$), where defects are likely to be interstitials. Such defects would only weakly affect the structure, and extensive TEM studies appear consistent with this identification, showing high intra-layer structural order even for dilute compositions (See Supplemental Information). From the data, the following empirical correlation can be discerned,
\begin{equation}
    {\rm sgn}(A_E) = -{\rm sgn}(A_B)\times{\rm sgn}(\delta), 
    \label{eq:nematic}
\end{equation}
where $A_E$~($A_B$) is the difference in resistivity $\Delta R_{\perp}/R_{//}$ observed in a switching (zAMR) experiment between vertical and horizontal electric pulses (applied magnetic fields). All compositions studied are consistent with this equation at all temperatures (Supplemental Information). As noted above, the parameter $A_B$ is sensitive only to the AFM order parameter, indifferent to the presence of the spin glass, whose sign measures the orientation of the N\'eel vector. The parameter $\delta$ in Eq. \ref{eq:nematic}, therefore plays the role of a $\mathcal{Z}_2$ nematic field, whose sign determines the equilibrium orientation of the N\'eel vector after an electrical pulse, as reflected by the sign of $A_E$. This suggests the local dynamics of the spin glass cause the N\'eel vector to either be rotated towards or away from the applied current pulse. The mechanism behind this must originate from differences in the microscopic spin texture of the spin glass in dilute and excess compositions; for example there may be differences in FM clustering that exchange-bias the response of the AFM\cite{lau2016spin} or perhaps the helicity of the spin texture about the defects changes\cite{gomonay2018antiferromagnetic}, distorting in orthogonal directions in the presence of an electrically driven spin accumulation. Whatever the magnetic texture dynamics responsible, the conclusion that spin is being imparted by the spin glass, with a direction determined by the microscopic nature of the spin glass, is inescapable.

The coupled response of the AFM and spin glass order parameters is unambiguous in the data, and is a significant departure from the usual mechanism driving spin-orbit torque based electrical switching of AFMs. The mechanism shares some commonality with FM/AFM heterostructures that leverage the spin angular momentum of the FM order and in some cases its magnons.\cite{meiklejohn1956new,Nogues1999,gomonay2018antiferromagnetic} In the present case, the collective behavior arises from the correlations between the defect species with its spinful environment. It would be interesting to study whether this collective  dynamics can additionally excite collective {\it modes } (the so-called Halperin-Salsow modes), but future experiments of non-local transport are necessary to confirm their existence.\cite{cornelissen2015long} Nevertheless, the collective behavior below $T_f$ opens up a new channel of spin transfer and maximizes the efficacy with which angular momentum is imparted to the AFM by the current pulse, making the spin glass an essential partner in the switching mechanism. It is worth noting as a concluding remark, that while spin glasses have been of extensive theoretical interest in condensed matter physics, they have been near absent in their application. The present work shows that while this mechanism is an uncommon way to leverage a spin glass to electrically switch an AFM, it need not be unique to \Fex{1/3+\delta}; correlated spin glasses appear generically in frustrated magnets,\cite{fischer1993spin} opening the field to candidate platforms that are in equal measure of applied and fundamental interest. 

\section {Methods}
Single crystals of \Fex{x} were synthesized using a chemical vapor transport technique. A polycrystalline precursor was prepared from iron, niobium, and sulfur in the ratio $x$:1:2 (Fe:Nb:S). The resulting polycrystalline product was then placed in an evacuated quartz ampoule with iodine as a transport agent (2.2 mg/cm$^3$), and put in the hot end of a two zone MTI furnace with temperature set points of 800$^\circ$C and 950$^\circ$C for a period of 7 days. High quality hexagonal crystals with diameters of several millimeters were obtained.
The iron intercalation values were confirmed by inductively coupled plasma optical emission spectroscopy (ICP-OES) using a Perkin Elmer Optima 7000 DV ICP-OES system and energy dispersive X-ray spectroscopy using an Oxford Instruments X-MaxN 50 $mm^{2}$ system. To perform the ICP-OES, the samples were first digested in hot 65\% nitric acid, which was subsequently treated with an excess of HF to ensure complete dissolution of niobium, and the solutions were subsequently diluted to appropriate concentrations.
Powder X-ray diffraction measurements were performed using a Rigaku Ultima-4 system with a Cu K-$\alpha$ radiation. 
Low field magnetization measurements were performed using a Quantum Design MPMS-3 system with a maximum applied magnetic field of 7 T.
Heat capacity was measured in a Quantum Design DynaCool PPMS system.
Electrical pulses were achieved using keithley 6221 current source.
NMR measurements were performed using the spin-echo technique, in the Condensed Matter NMR lab at NHMFL, using a home-built NMR spectrometer with quadrature detection. The magnetic field was varied between 6 T and 10 T at various temperatures from 4.2 K to 100 K.
For the thermodynamic, zAMR and NMR measurements bulk single crystals were used. 
The switching devices required fabrication of bulk crystals into defined micro-structures using a Focused Ion Beam microscope as described in our previous study.\cite{nair2019electrical}
HAADF-STEM images were recorded using the TEAM I at the Molecular Foundry: an aberration-corrected STEM (Thermo-Fischer Titan Cubed 80-300kV) fitted with a high-brightness field-emission gun (X-FEG), a CEOS DCOR probe corrector operated at 300 kV. The beam convergence angle was 30 mrad, and thus yields a probe size of less than 0.10 nm under STEM mode.
\section{Acknowledgements} This work was supported as part of the Center for Novel Pathways to Quantum Coherence in Materials, an Energy Frontier Research Center funded by the US Department of Energy, Office of Science, Basic Energy Sciences. J.G.A., E.M. and N.L.N. was funded in part by the Gordon and Betty Moore Foundation’s EPiQS Initiative, Grant GBMF9067 to J.G.A. FIB device fabrication was performed at the National Center for Electron Microscopy at the Molecular Foundry. Work at the Molecular Foundry was supported by the Office of Science, Office of Basic Energy Sciences, of the US Department of Energy (contract no. DE-AC02-05CH11231).
A portion of this work was performed at the National High Magnetic Field Laboratory, which is supported by the National Science Foundation Cooperative Agreement No. DMR-1644779 and the State of Florida.
\section{Author Contribution} E.M., S.C.H., S.D. and C.J. performed crystal synthesis and magnetization measurements. E.M. performed heat capacity and EDS measurements. E.M., N.L.N. and S.C.H. fabricated the FIB devices and performed transport measurements. Y.L.T. and P.E. performed transmission electron microscopy measurements and analysis. A.M., S.K.R. and A.P.R. performed NMR measurements. E.M. and J.G.A. performed data analysis and wrote the manuscript with input from all coauthors.  
\section{Competing Interests} A patent has been filed by Lawrence Berkeley National Laboratory on behalf of J.G.A., E.M., N.L.N., C.J. and S.D. pertaining to the use of Fe$_{1/3}$NbS2 and related intercalated TMD compounds in AFM spintronic devices as described in this manuscript under US Patent Application Ser. No. 62/878,438.
\section{Correspondence} Correspondence and requests for materials
should be addressed to E.M.~(email: eranmaniv@berkeley.edu) or J.G.A.~(email: analytis@berkeley.edu).

%

\end{document}